\documentclass[graybox]{svmult}

\usepackage{type1cm}        % activate if the above 3 fonts are
\usepackage{makeidx}         % allows index generation
\usepackage{graphicx}        % standard LaTeX graphics tool
\usepackage{multicol}        % used for the two-column index
\usepackage[bottom]{footmisc}% places footnotes at page bottom

\usepackage{newtxtext}       % 
\usepackage[varvw]{newtxmath}       % selects Times Roman as basic font

\usepackage{subfigure}

\makeindex             % used for the subject index
\begin{document}

\title*{Flow Cytometry Quantification of Transient Transfections in Mammalian Cells}
\author{Jacob Beal}
\institute{Raytheon BBN, 10 Moulton Street, Cambridge, MA, USA, \email{jakebeal@ieee.org}}
\maketitle

% abstract is up to 200 words
\abstract{
Flow cytometry is a powerful quantitative assay supporting high-throughput collection of single-cell data with a high dynamic range.
For flow cytometry to yield reproducible data with a quantitative relationship to the underlying biology, however, requires that 
1) appropriate process controls are collected along with experimental samples, 
2) these process controls are used for unit calibration and quality control, and 
3) data is analyzed using appropriate statistics.
To this end, this article describes methods for quantitative flow cytometry through addition of process controls and analyses, thereby enabling better development, modeling, and debugging  of engineered biological organisms.
The methods described here have specifically been developed in the context of transient transfections in mammalian cells, but may in many cases be adaptable to other categories of transfection and other types of cells.}

\section{Introduction}
\label{sec:introduction}

Flow cytometry is a long-established method for assaying the optical properties of large numbers of single cells, in use for more than half a century~\cite{fulwyler1965electronic, kamentsky1973cytology}.
Much of the development and use of flow cytometry has been in clinical applications focused on categorization of cells into classes, such as immunoassays and diagnosis of blood diseases.
In recent years, however, flow cytometers have also been used to great effect in synthetic biology for measuring the properties of individual cells, e.g., measuring the expression level of a gene by estimating the number of molecules of a proxy fluorescent reporter.

In a typical flow cytometer, a stream of cells are passed in front of one or more lasers, with an arrangement of optical filters and detectors collecting information on the light scattering and fluorescence properties of each cell-like particle as it passes.
Flow cytometer optical detectors often have high sensitivity and range, with signals distinguishable on the order of a $10^3$ to $10^6$ ratio between the highest and lowest distinguishable values. 
Data is typically collected from on the order of $10^3$ to $10^5$ particles per sample over the time span from seconds to minutes, and automated samplers allow high-throughput processing of dozens to hundreds of samples from microplates.

\begin{figure}[h]
\centering
\includegraphics[width=0.9\textwidth]{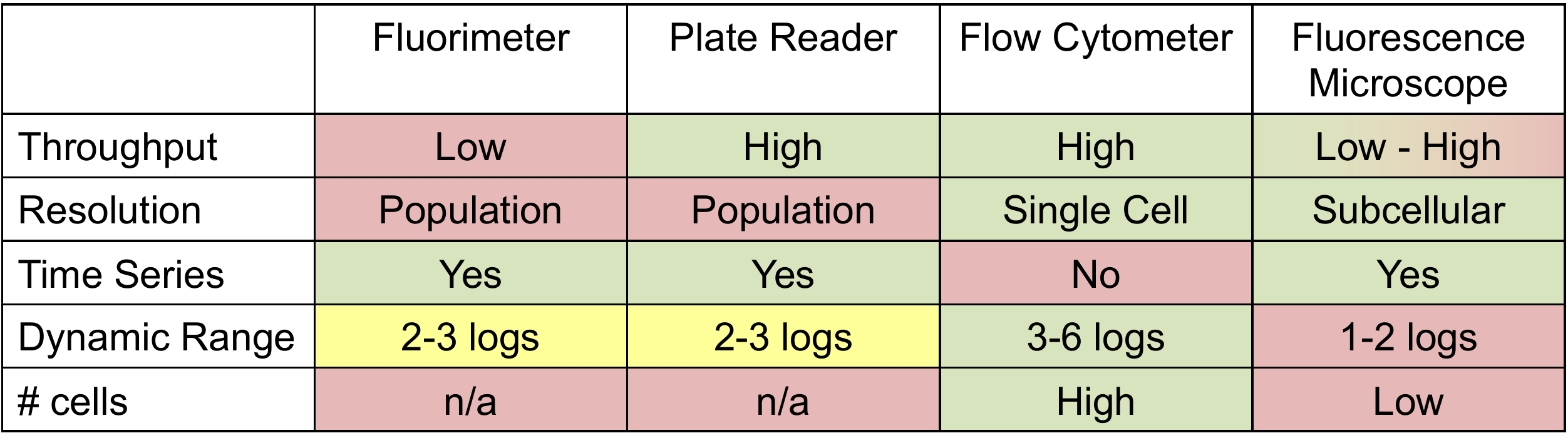}
\caption{Flow cytometry provides high-dynamic range data from large numbers of single cells at a single time point, complementary to the properties of other common fluorescence-based assays.}
\label{f:assays}
\end{figure}

Flow cytometers can thus be prodigious sources of single-cell data with a high dynamic range, and form an important part of the synthetic biology engineering toolkit.
Compared to other broadly accessible fluorescence assays (Figure~\ref{f:assays}), flow cytometry shines in number of cells assayed and sensitivity (i.e., dynamic range).
Because cells must be suspended in sheath fluid and passed through the instrument, however, the assay is typically disruptive to cell state and thus generally not suitable for collecting time series data.
Plate readers provide a useful complementary capability for flow cytometry, since they are effective at collecting time series data and can be calibrated to equivalent units~\cite{beal2021comparative,beal2020robust}.
Likewise, microscopy provides the complementary capabilities of subcellular resolution and time series data. 
Indeed, imaging flow cytometry can fuse microscopy and flow cytometry by adding a microscopy stage to a flow cytometer~\cite{barteneva2012imaging}, though it is still disruptive to cell states.

The relationship between the properties of a cell and the optical signal measured in a channel of a flow cytometer, however, is highly sensitive to specifics of the instrument's configuration, current settings, and level of optical wear, as well as to choice of fluorescent reporter, cell state, and interference from other measurement channels~\cite{Roederer2001,Roederer2002,gaigalas2005quantitating}. 
As a consequence, interpreting flow cytometry data to produce precise, reproducible, and biologically relevant estimates of molecule counts requires careful use of process controls linked to appropriate quality control and analytical methods~\cite{beal2022meeting}.
When appropriate process control, calibration, and analysis protocols are applied, however, flow cytometry can produce reliably produce reproducible measurements with a quantitative relationship to the underlying biology.

This article aims to increase the accessibility of quantitative flow cytometry by summarizing methods for:
\begin{enumerate}
\item enhancing a flow cytometry study by adding process controls to support calibration and quantification, 
\item using data from these process controls for unit calibration and quality control, and 
\item analysis of the resulting calibrated data using appropriate statistics.
\end{enumerate}

More specifically, this article focuses on methods that were developed in the context of transient transfections in mammalian cells over the course of a number of studies that used these methods to develop, model, refine, and predict the behavior of a variety of genetic devices~\cite{beal2012method,DavidsohnBeal14,kiani2014crispr,kiani2015cas9,beal2015model,wagner2018small,siciliano2018engineering,weinberg2019high,kiwimagi2021quantitative}.
Note that these methods are modular with respect to most aspects of cell type, transfection, and culturing, so the actual preparation of the experimental samples will be largely abstracted, while the methods focus on the process controls necessary for effective calibration, and the analysis used to make use of those process controls.
Likewise, many aspects of the methods should be adaptable to non-transient transfections and to non-mammalian cells, and to this end the methods will explicitly identify the points where enabling assumptions are made based on mammalian transient transfection in order to better understand limits and enable adaptation to other contexts.

\section{Materials}
\label{sec:materials}

For calibration and quality control on a flow cytometry study, two classes of calibrants are needed: calibrant reagents and cellular calibrants.

Calibrant reagents are materials with known optical properties that are used to compute a unit conversion factor between the arbitrary unit values output by a flow cytometer and the physical properties of size and molecule count.
Failures in calibrant reagents are typically uncorrelated with failures in the preparation of the experimental samples to be assayed, so these calibrants serve as a process control testing for correct operation of the flow cytometer.

Cellular calibrants, on the other hand, are positive and negative controls that establish the expected dynamic range of cellular behaviors in each fluorescent channel and the relationships between measurements in different channels.
Failures in cellular calibrants are typically correlated with failures in the preparation of experimental samples, so these calibrants serve as a process control testing for correct execution of the preparation of experimental samples.

\subsection{Calibrant Reagents}

\begin{itemize}
\item {\bf SpheroTech Ultra Rainbow Quantitative Particle Kit (URQP-38-6K)} 
 \begin{itemize}
 \item This kit provides beads with six different intensities of fluorescence and NIST-certified Equivalent Reference Fluorophore (ERF) value that map each intensity to an equivalent number of fluorescent molecules on four channels: fluorescein, Nile Red, APC and Coumarin 30. Calibrant should be stored, handled, and dispensed according to the manufacturer directions. 
 One kit provides sufficient materials for many assays.
 \item These beads are used to compute a linear unit conversion factor that maps fluorescence channel values to molecules of equivalent fluorophore. The specific units depend on the channel, e.g., Molecules of Equivalent FLuorescein (MEFL).
 \item Alternate beads may be substituted, but must have multiple separable intensities and certified ERF values, one set of which must be for fluorescein.
 \end{itemize}

\item {\bf SpheroTech Polystyrene Particle Size Standard Kit (PPS-6K)} 
 \begin{itemize}
 \item This kit provides beads with six different diameters. Calibrant should be stored, handled, and dispensed according to the manufacturer directions.
 One kit provides sufficient materials for many assays.
 \item These beads are used to compute a non-linear unit conversion function that maps forward scatter area (FSC-A) values to estimated diameter in units of equivalent micron particle diameter (E$\mu$m).
 \item Alternate beads may be substituted, but must be monodisperse and must have multiple separable sizes. 
 \item Certified size beads are not currently available, but should be preferred if they become available in the future.
 \end{itemize}
\end{itemize}

\subsection{Cellular Calibrants}

\begin{figure}
\centering
\includegraphics[width=1.0\textwidth]{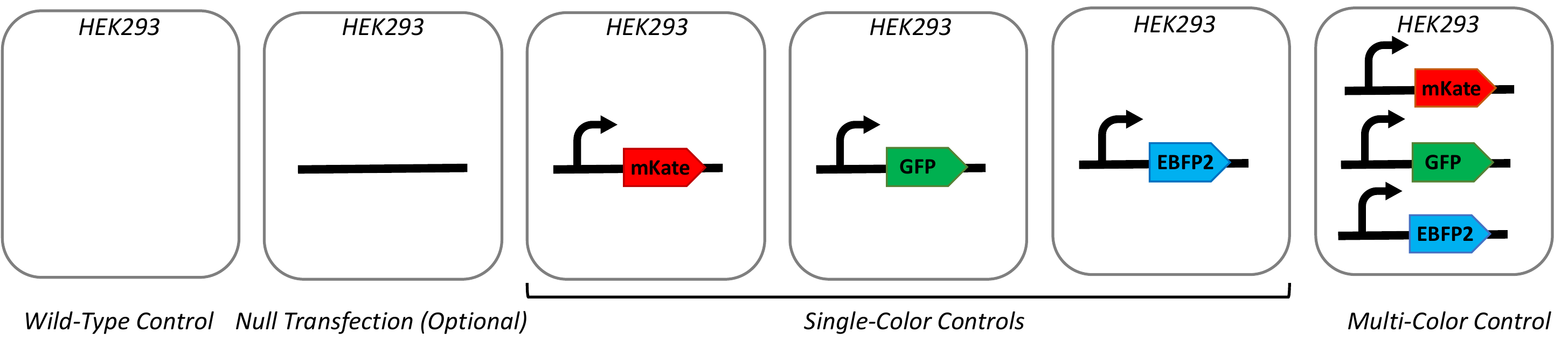}
\caption{Example genetic constructs for cellular controls, for an experiment with red and green fluorescence, adding blue fluorescence as a transfection marker, plus the optional null transfection control (see Section~\ref{sec:nt}).}
\label{f:cellcontrols}
\end{figure}

\begin{itemize}
\item {\bf Wild Type Negative Control} 
 \begin{itemize}
 \item For each strain of cell used, at least one replicate with a sample of wild-type cells, i.e., not transformed with any construct. For example, if CHO and HEK293 cells are included in the experiment, there should be at least one sample of wild-type CHO and at least one sample of wild-type HEK293.
 \item The wild type control is used to estimate autofluorescence and to assess strain response to culturing and/or other sample preparation activities.
 \item The wild type control should be cultured and/or otherwise prepared alongside the experimental samples, but not given any other experimental treatment.
 \end{itemize}

\item {\bf Single Color Controls}
 \begin{itemize}
 \item For each fluorescent reporter used, at least one replicate transiently transfected with a construct for strong constitutive expression of the fluorescent reporter (and no other genetic function in its design). 
 \item Transient transfection is preferable to integration, as it is both faster and produces a wider range of expression levels, including some that are much stronger.
 \item The single color controls are used estimate spectral overlap between channels and to assess strain response to culturing.
 \item Fluorescent reporters should be selected such that no pair of channels have more than a 10\% spectral overlap. This typically allows three fluorescent reporters (red/green/blue), or four if deep red / infrared fluorescence is used.
 \item The single color controls should be cultured and/or otherwise prepared alongside the experimental samples, but not given any other experimental treatment.
 \end{itemize}
 
 \item {\bf Multi-Color Control}
 \begin{itemize}
 \item At least one replicate transiently co-transfected with equal doses of single color control constructs for all  of the fluorescent reporters used, where the constructs differ only in their fluorescent reporter (i.e., the vector and regulatory regions are identical) and have identical dosages in the co-transfection.
 \item The multi-color control is used to estimate unit conversions from fluorescent channels, such that the expression levels of different fluorescent reporters can be directly compared in the same molecular units. 
If only reference fluorophores are used, then ERF values can be compared directly, but reference fluorophores are typically small-molecule dyes rather than biologically-expressed reporter constructs.
With biologically expressed reporters, ERF values for different channels are generally incommensurate due to the differences in relationship between fluorescent reporter and reference fluorophore. 
For example, the relationship between fluorescein and GFPmut3 in the FITC-A channel is expected to be different than the relationship between Coumarin 30 and EBFP2 in the Pacific Blue-A channel.

 \item The recommended unit to convert to is Molecules of Equivalent FLuorescein (MEFL), as fluorescein channels are common and often have similar spectral properties, and as fluorescein has similar spectral properties to those of many green fluorescent reporters.
 \item It is preferable that the multicolor control have at least three colors, as this will allow better segmentation into subpopulations for analysis.
 \item The multi-color control should be cultured and/or otherwise prepared alongside the experimental samples, but not given any other experimental treatment.
 \item This control may be omitted if the only experimental fluorescent reporter is measured in the fluorescein channel (e.g., GFP, YFP). If there is no experimental reporter measured in the fluorescein channel, then a strong constitutive expression construct for such a reporter should be added to the multi-color control and also as an additional single color control.
 \item This control depends on the relatively low sensitivity of mammalian promoter transcription rates to the contents of the coding sequences being transcribed, such that the transcriptional activity of each co-transfected construct can be expected to be equivalent. This control also depends on co-transfection introducing a large number of plasmids to each cell (typical of many protocols for mammalian transient transfection, e.g., lipofection), which means that strongly transfected cells should have a ratio between plasmids tightly distributed around 1:1.  These assumptions often do not hold for other types of cells or transformations.
 \end{itemize}
\end{itemize}

\subsection{Experimental Samples}

\begin{itemize}
\item Experimental samples should be prepared according to the experimenter's study design.
\item The single color control for a non-conflicting fluorescent reporter should be co-transfected along with the experimental construct or constructs. During analysis, this fluorescence channel will be used as a transfection marker, enabling parametric analysis of behavior with respect to transfection copy count.
\item Experimental samples should be cultured along with the cellular calibrants and should be measured at the same time as both the calibrant reagents and cellular calibrants, in order to maximize the likelihood that issues with the protocol or instrument can be detected. Note that single-color and multi-color controls may simultaneously server as experimental samples, if appropriate for the experiment design.
\end{itemize}

\section{Methods}
\label{sec:methods}

The analytical methods below are described with sufficient detail to allow them to be understood and implemented by the reader.
Note, however, that application of these analysis methods can be simplified by using the TASBE Flow Analytics Matlab software package~\cite{beal2019tasbe}, an existing implementation of methods and workflows to automate all of the steps described in this section, as well as to support the debugging steps listed below in Section~\ref{sec:notes}.
Many of these steps are also automatable using existing Python implementations in  FlowCal~\cite{castillo2016flowcal} or CytoFlow~\cite{teague2022cytoflow}.

\subsection{Data Collection}
\label{s:capture}

\begin{figure}
\centering
\includegraphics[width=0.9\textwidth]{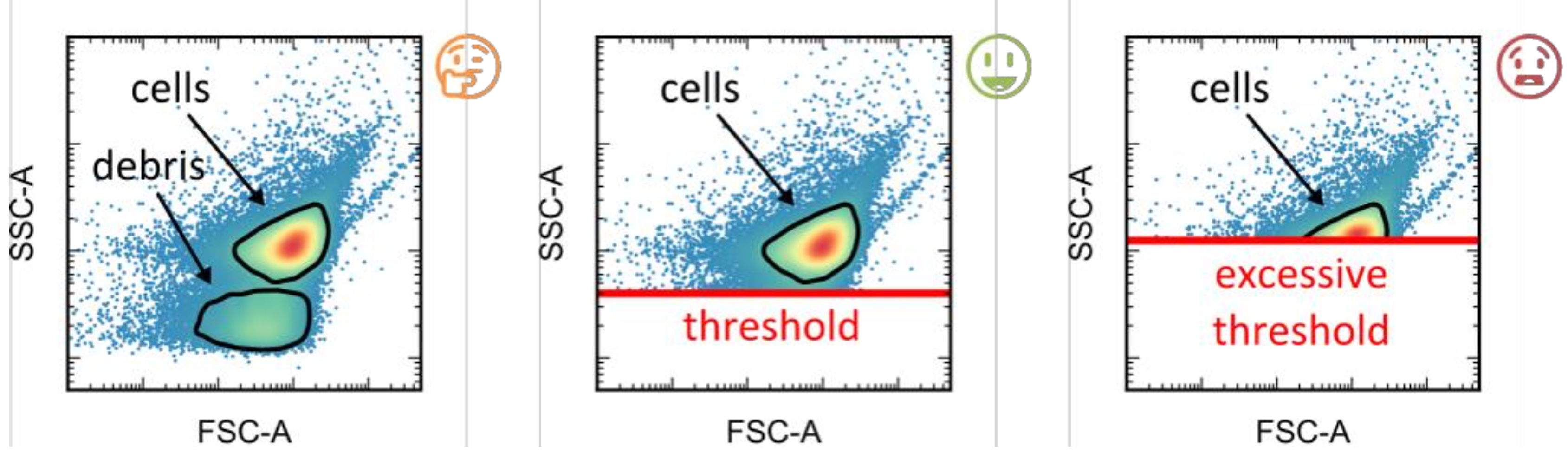}
\caption{Examples of data collection gating with size thresholds that are too permissive (left), appropriate (center), and too restrictive (right), adapted from protocol in~\cite{beal2022meetingBioRxiv}}
\label{f:thresholding}
\end{figure}

\begin{itemize}
\item Set a conservative forward scatter (particle size) trigger, such that the number of small non-cell particle events is minimized while not removing smaller cells (Figure~\ref{f:thresholding}).
 \begin{itemize}
 \item The minimum size threshold trigger is needed because flow cytometers typically have  limits in the rate at which events can be recorded, and if the instrument's recording capability is saturated then many cell measurements will be lost.
 \item Other instrument gating should not be used, as it is generally better to use post-facto filtering to ignore unneeded events, rather than to discard cell data that may later prove valuable. 
 \end{itemize}

\item Record all forward scatter (FSC) and side scatter (SSC) channels, and the area channel for each fluorescent reporter.
 \begin{itemize}
 \item Forward scatter and size scatter will be used to gate out non-cell particles and multi-cell clumps, and thus the height and width channels, if available, are useful in addition to the area channels.
 \item In fluorescent channels only area is needed, as that is the channel that will be used to estimate equivalent molecule counts.
\item Ensure that the instrument is putting laser and filter information in FCS files, and that the values for these are correct. This is valuable for ensuring that instrument settings are kept consistent between samples.
 \end{itemize}

\item Gather between 50,000 and 100,000 events per sample.
 \begin{itemize}
 \item In a transient transfection, cells generally receive a widely varied number of copies of the plasmid or co-transfected set of plasmids. Collecting a larger number of events allows parametric analysis in which events are divided into subpopulations by transfection marker (see Section~\ref{s:analysis}).
 \item If subpopulation analysis is not needed needed (e.g., for integrated constructs), the number of events can be reduced to 10,000 to 30,000.
 \end{itemize}
\end{itemize}

\subsection{Log-Scale Statistical Analyses}
\label{s:analysis}

\begin{figure}
\centering
\includegraphics[width=0.8\textwidth]{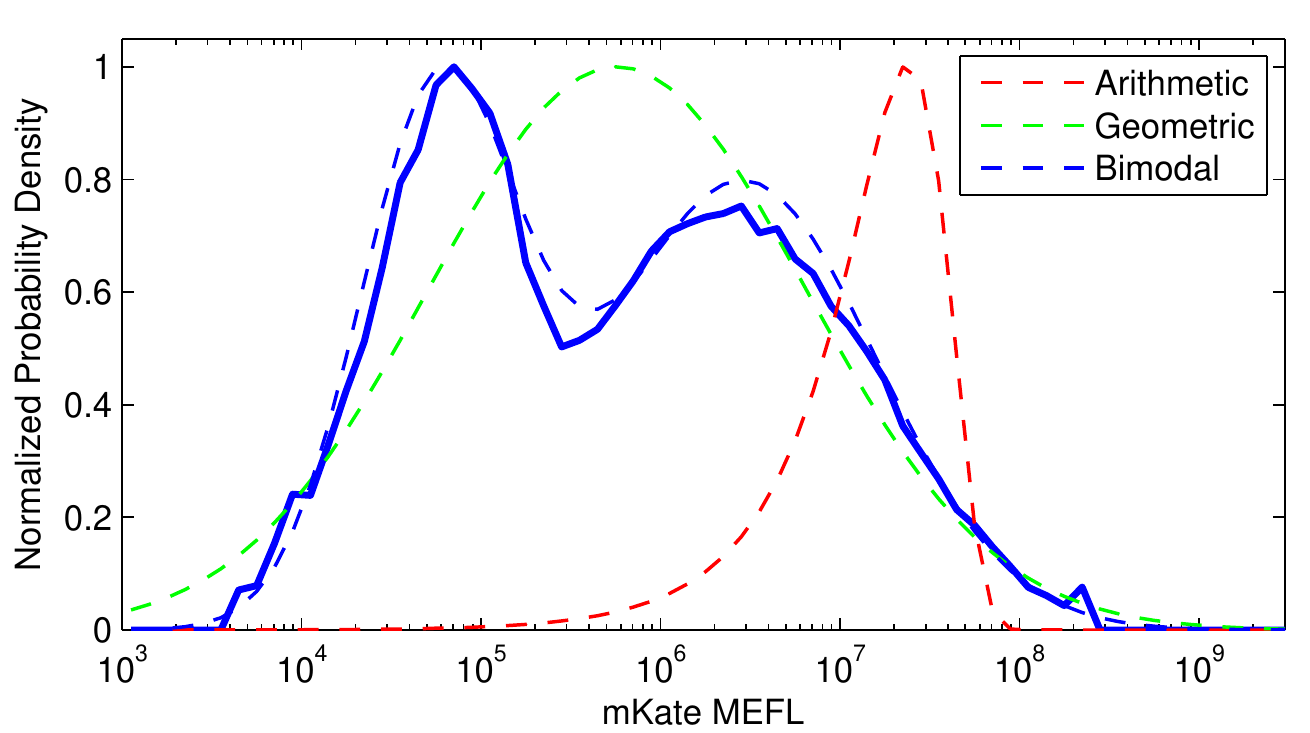}
\caption{Cell behavior often shows distributions that are better suited to geometric statistics on the log scale, rather than arithmetic statistics on the linear scale. In many cases, fits with bi-modal or multimodal log-normal distributions are appropriate. Figure adapted from~\cite{beal2017lognormal}.}
\label{f:geometric}
\end{figure}

All of the subsequent stages of this protocol are analytical, and most depend on log-scale statistical analyses of data, so here we will discuss the reasons for these analysis and provide their formulations for those who may be unfamiliar.

Log-scale analysis is motivated by the fact that strong gene expression typically has a log-normal distribution~\cite{beal2017lognormal}.
Transient transfections in mammalian cells typically deliver many copies of a genetic construct to each cell, and fluorescent reporter signals are typically engineered to be as strong as an experimenter can obtain, both of which mean that strong overall expression is the typical case.
For this reason, cell behavior often shows distributions that are better suited to geometric statistics on the log scale, rather than arithmetic statistics on the linear scale, as illustrated in the example in Figure~\ref{f:geometric}
(other statistical alternatives are discussed in Section~\ref{s:alt_stats}).

Geometric mean and geometric standard deviation may be computed by computing the aithmetic mean and standard deviation of a variable that has been transformed onto the log scale:
\begin{itemize}
\item Geometric mean of variable $X$:  $$\mu_g(X) = 10^{\mu(log_{10}(X))}$$ 
\item Geometric standard deviation of variable $X$:  $$\sigma_g(X) = 10^{\sigma(log_{10}(X))}$$ 
\end{itemize}

Comparison of these statistics should also be discussed with respect to geometric (fold) differences.
For example, a decrease from 4000 MEFL to 1000 MEFL or 250 MEFL should be interpreted as a 4-fold or 16-fold decrease, respectively, and not 75\% or 94\%.

Histograms should likewise generally be computed on the log scale.
For purposes of the analyses discussed here, histograms should be computed using a granularity of 10 bins per decade: for example, in the range between 100 and 1000, there is a bin for values from 100 to 125.9, from 125.9 to 158.5, etc., until the last bin hold values from 794.3 to 1000.
This may be computed by taking a histogram of data transformed onto the log scale.
This granularity matches well with typical flow cytometer precision and the number of data points recommended for capture in Section~\ref{s:capture}.
If bin-to-bin variability is too high, however, the granularity may be reduced, e.g., from 10 bins per decade to 5 bins per decade.

Parametric analysis combines histograms on the log scale with geometric statistics.
Parametric analysis is computed by grouping flow cytometry events into log-scale bins with respect to a selected fluorescent channel, at a granularity of 10 bins per decade.
Then, for each bin, the geometric mean and standard deviation of other fluorescent channels are calculated over all of the events in the bin.
Bins with less than a minimum count of 10 events are excluded from the parametric results, due to the uncertainty of values computed from small numbers of data points.
The result is a function capturing the relationship between the selected channel and the other fluorescence channels.
Examples of its use are measuring the spillover fluorescence in a spectrally overlapping channel as a function of fluorescence in the intended channel (Section~\ref{s:compensation}), or the value of an experimental variable as a function of transfection marker expression (Section~\ref{s:expt_analysis}).

\subsection{Bead-based Fluorescence Channel Calibration}
\label{s:colorbeads}

\begin{figure}
\centering
\subfigure[MEFL bead peaks]{\includegraphics[width=0.45\textwidth]{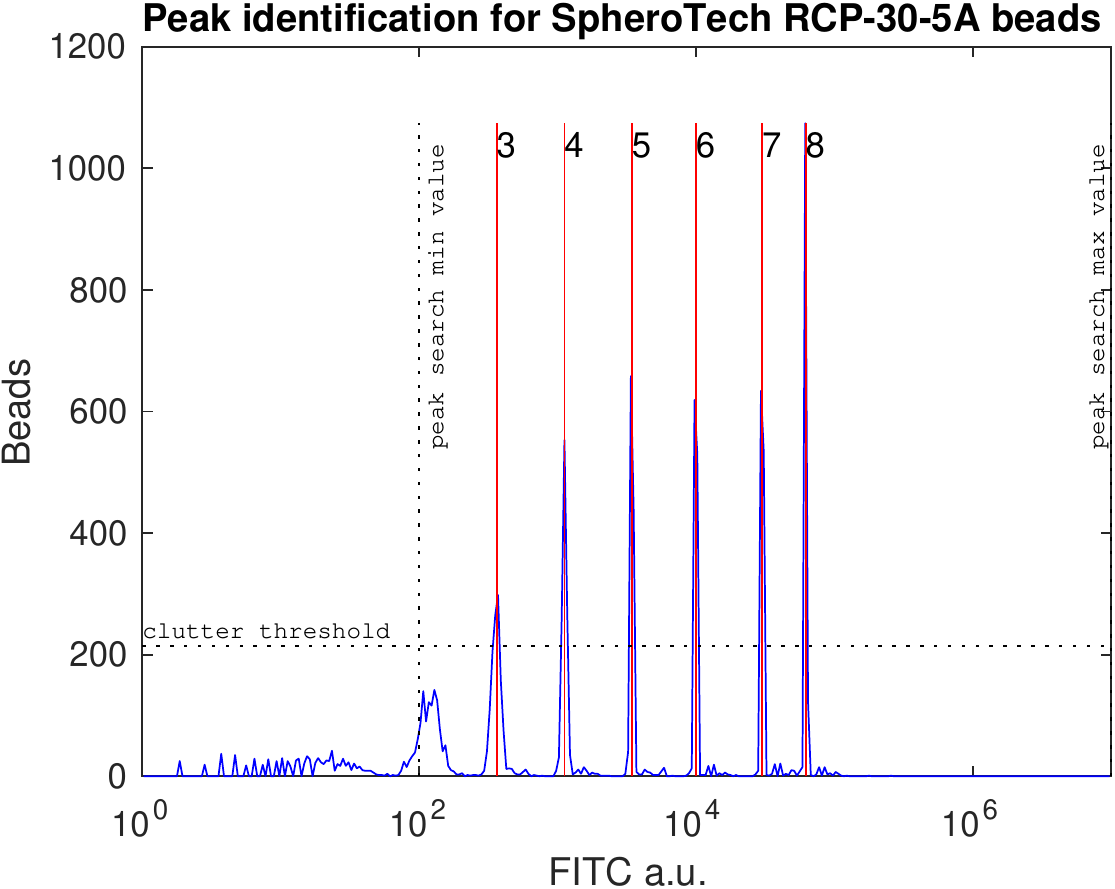}\label{f:color_peaks}}
\subfigure[MEFL bead fit]{\includegraphics[width=0.45\textwidth]{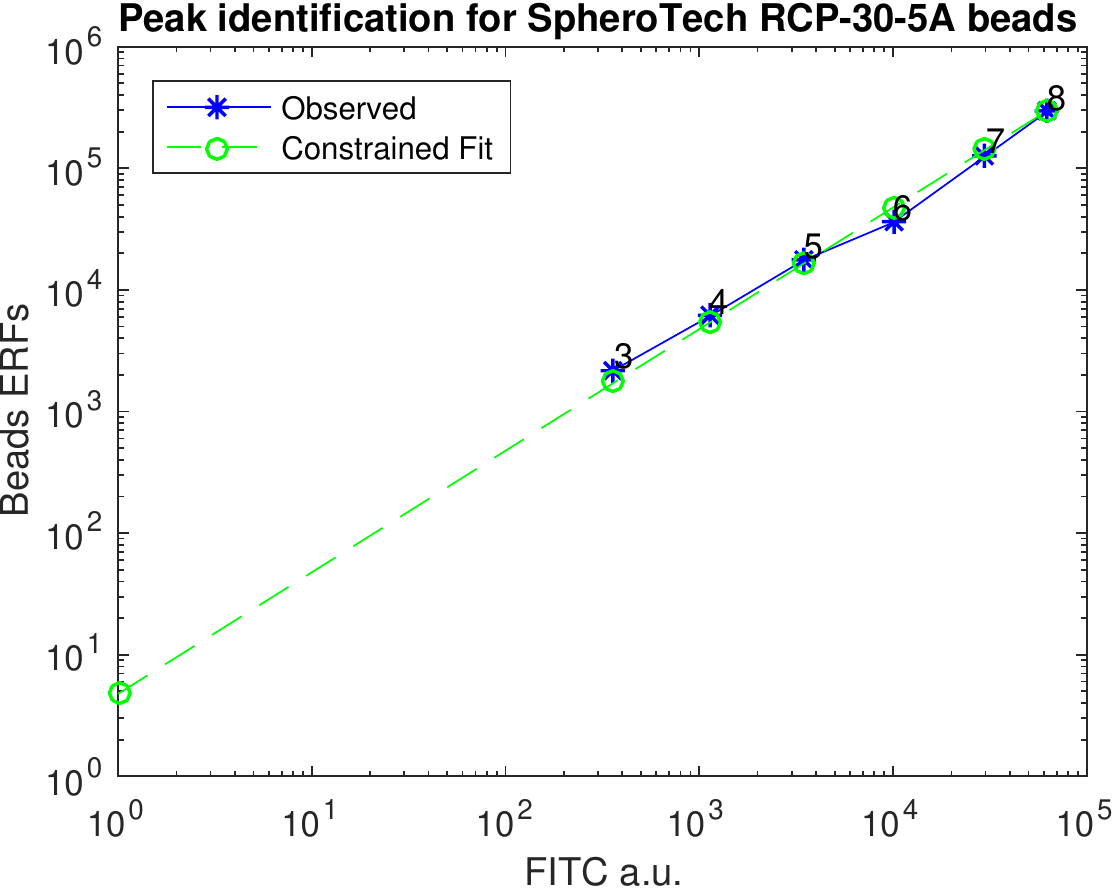}\label{f:color_fit}}
\caption{Example of color bead analysis, showing (a) peak identification for the fluorescein channel against an alternate calibration bead, SpheroTech RCP-30-5A. This is an 8-peak calibrant, and in this case only the top 6 peaks are being used. Note the small doublet peaks slightly above each identified peak. (b) Scaling fit for unit conversion function mapping FITC area channel arbitrary units to Molecules of Equivalent FLuorescein (MEFL). Figure adapted from~\cite{tasbe_tutorial}.}
\label{f:color}
\end{figure}

For each channel for a fluorescent reporter, compute the unit scaling between arbitrary units on the fluorescence channel and units of Equivalent Reference Fluorophore (ERF) for that channel:
\begin{itemize}
\item Compute a log-scale histogram of the channel values for the rainbow bead sample.
\item Identify the peaks in the channel and compute the geometric mean for each peak region, as shown in the example in Figure~\ref{f:color_peaks}.
  \begin{itemize}
  \item Channel range (e.g., voltage) should be set based on the cells to be measured, not the beads. It is acceptable for some beads to be saturated high, lost to low noise, or otherwise not have usable peaks. For purposes of fit quality, however, it is preferable to have at least three usable peaks.
  \item Low-valued peaks, typically below around $10^2$ arbitrary units, will often be blurred due to instrument noise or other issues. It is better to exclude these peaks and use only the better-defined peaks at higher values.
  \item Bead doublets will often form a small peak with a value twice that of the primary peak for each expected bead intensity. Sometimes, a third triplet peak with a three-times value will also be distinguishable. Data from these doublet and triplet peaks should be excluded from peak calculations.
  \end{itemize}
\item Perform a constrained linear fit (slope=1) on the log scale between peak values and manufacturer supplied peak ERF values, as shown in the example in Figure~\ref{f:color_fit}.
  \begin{itemize}
  \item Mean fit error should be under 10\%. Higher error typically indicates that peaks have not been correctly identified: see Section~\ref{s:bead_problems} for likely issues.
  \end{itemize}
\end{itemize}

Remember that the ERF values are not actually number of molecules of the fluorescent reporter, but the amount of fluorescence that would be produced by that number of molecules of the reference fluorophore.
With good spectral matching between the fluorescent reporter, reference fluorophore, and measurement channel, however, the impact of this difference is likely to be less than the magnitude of error from other sources~\cite{beal2022meeting}.
Fluorescent reporter molecules also may not properly mature to give full fluorescence due to biological issues such as incomplete folding, lack of oxygen, pH range, etc.

When channel conversion is used (Section~\ref{s:conversion}), only the Molecules of Equivalent FLuorescein (MEFL) scaling factor will actually be used. Nevertheless, unit scaling should be computed for all fluorescence channels, as the ability to compute a reliable unit scaling all channels is a process control indicating correct instrument operation.

\subsection{Bead-based Particle Size Calibration}
\label{s:sizebeads}

\begin{figure}
\centering
\subfigure[Size bead peaks]{\includegraphics[width=0.45\textwidth]{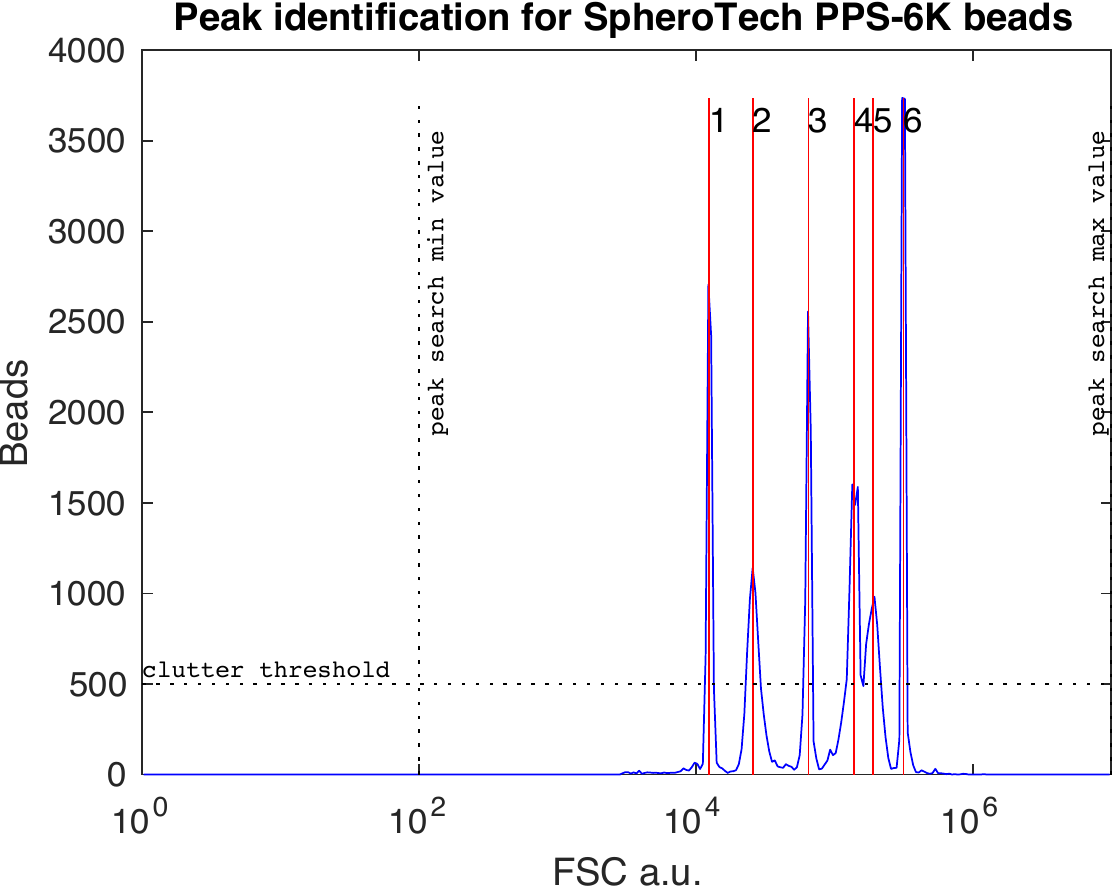}\label{f:size_peaks}}
\subfigure[Size bead fit]{\includegraphics[width=0.45\textwidth]{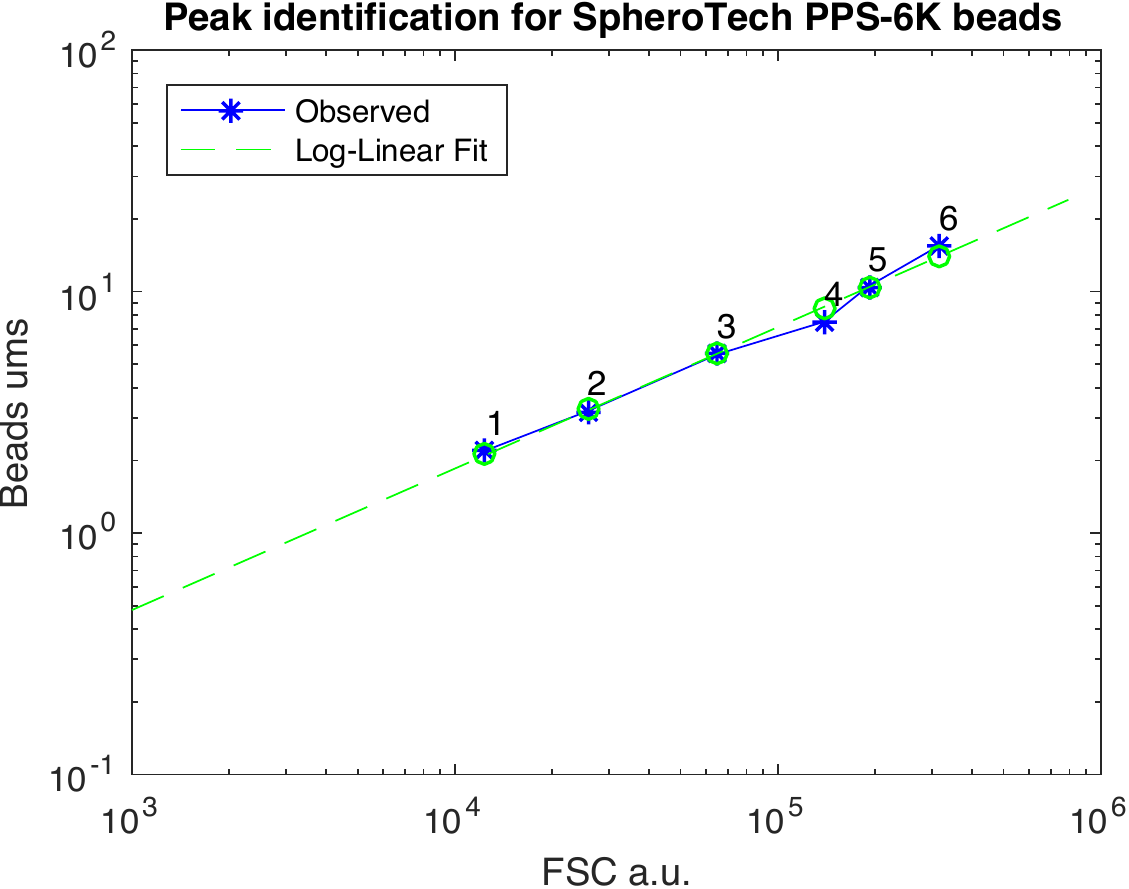}\label{f:size_fit}}
\caption{Example of size bead analysis, showing (a) peak identification, and (b) log-log fit for unit conversion function mapping forward scatter area (FCS-A) channel arbitrary units to equivalent micro-meter (E$\mu$m) diameter. Note that the fit function does not a have a 1:1 slope in the log scale, meaning that it is a not a linear scaling relationship.
Figure adapted from~\cite{tasbe_tutorial}.}
\label{f:size}
\end{figure}

\begin{itemize}
\item Compute a log-scale histogram of the forward scatter area (FSC-A) channel values for the size bead sample.
\item Identify the peaks in the channel and compute the geometric mean for each peak region, as shown in the example in Figure~\ref{f:size_peaks}.
  \begin{itemize}
  \item Channel range (e.g., voltage) should be set based on the cells to be measured, not the beads. Due to the smaller range of typical cell sizes, however, it is typically possible to set FSC-A range to be appropriate for cell measurement and also to keep all beads in a range where peaks can be clearly distinguished.
  \item If the channel cannot be adjusted to make all bead peaks distinguishable, it is acceptable for some beads to be saturated high, lost to low noise, or otherwise not have usable peaks. For purposes of fit quality, however, it is preferable to have at least three usable peaks. 
  \item Likewise, if there are low-valued peaks, typically below around $10^2$ arbitrary units, that are blurred due to instrument noise or other issues. It is better to exclude these peaks and use only the better-defined peaks at higher values. 
  \item Note that at present there is generally a higher degree of variability in size beads than rainbow beads due to differences in the available calibrant materials. Doublet and triplet peaks are typically less of an issue with size beads, however, due to the lower amount of separation between peaks.
  \end{itemize}
\item Compute the unit scaling factor from FSC-A arbitrary units to equivalent micro-meter (E$\mu$m) diameter by performing a linear fit on the log scale between peak values and manufacturer supplied peak diameter values, as shown in the example in Figure~\ref{f:size_fit}.
  \begin{itemize}
  \item Unlike with color beads, this conversion should not have a unit slope, as there is a non-linear relationship between diameter and forward scatter~\cite{koch96FSC,chandler11FSC}. Unit transformation will thus be a non-linear function.
  \item Mean fit error should be under 10\%. Higher error typically indicates that peaks have not been correctly identified: see Section~\ref{s:bead_problems} for likely issues.
  \end{itemize}
\end{itemize}

\subsection{Cell Gating}
\label{s:gating}

\begin{figure}
\centering
\includegraphics[width=0.6\textwidth]{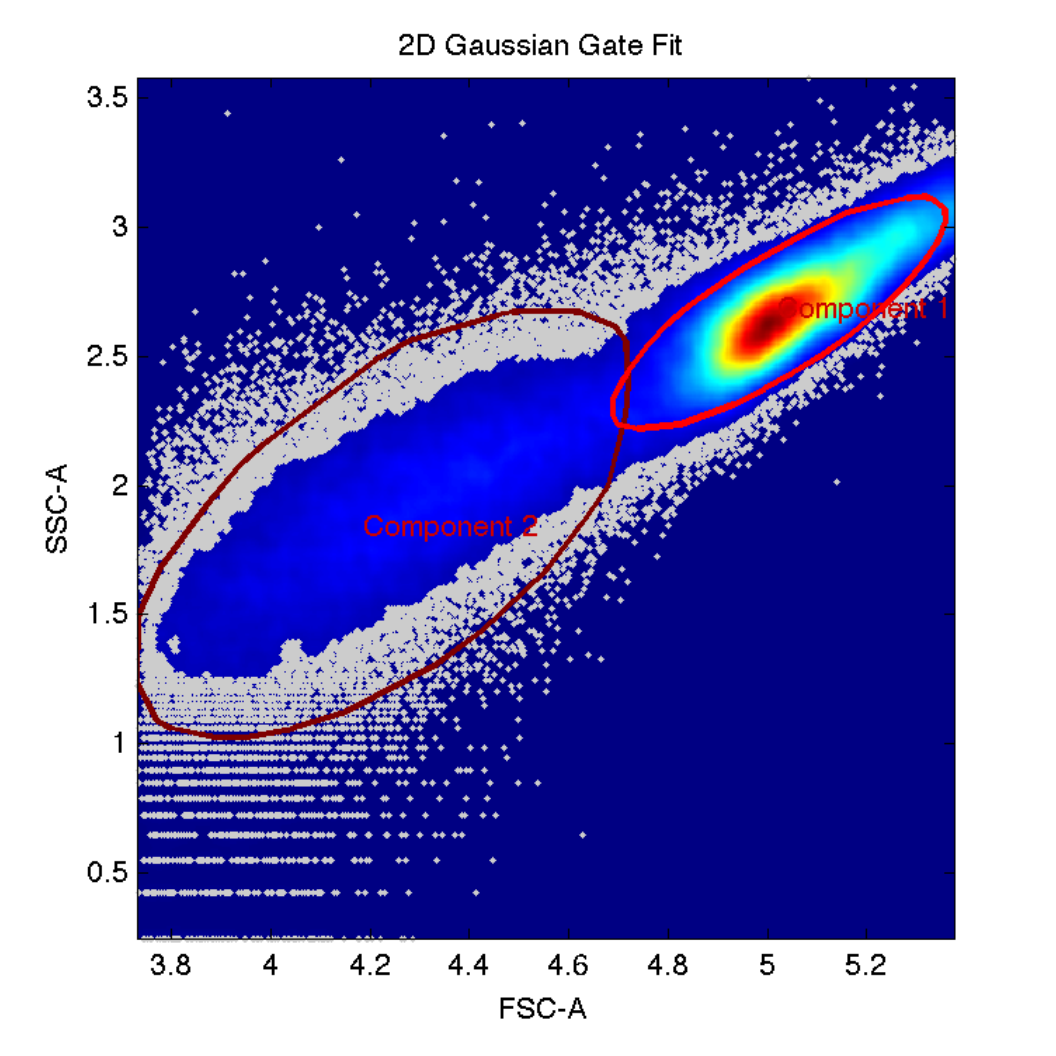}
\caption{Example of gate to select cell events via a Gaussian mixture model (GMM), in this case a two-component GMM built against forward and side scatter area. The smaller, bright red, component selects events likely to be from cells.
Figure adapted from~\cite{tasbe_tutorial}.}
\label{fig:gating}
\end{figure}

\begin{itemize}
\item Load events from the negative control (or null transfection, if one was used per Section~\ref{sec:nt})
\item Compute a two-component n-dimentional Gaussian mixture model (GMM) fit jointly on all the forward and side scatter channels.
  \begin{itemize}
  \item Forward scatter area (FSC-A) and side scatter area (SSC-A) channels are often sufficient for this purpose. 
  \item Including height and width channels can help further refine cell selection, particularly with respect to excluding doublet events.
  \end{itemize}
\item To gate for cells, select all events that are within two standard deviations of the mean of the GMM component with the smallest standard deviation, and also not closer to the mean of any other component. These are candidate cell events.
\item Plot a two-dimensional density map of all events, with a convex hull around the selected events.
  \begin{itemize}
  \item Optionally, also plot the convex hulls of the events that would be selected for each other component.
  \item Figure~\ref{fig:gating} shows an example of selecting the cell component for a two-component GMM.
  \item Compare selected component with size beads to determine is estimated diameter of cell-like events is consistent with experimental expectations.
  \item If the distribution of events is such that cell-like events are not well-identified by two standard deviations from the smaller standard deviation component of a two-component GMM, then select a different component, adjust the number of components, and/or adjust the number of standard deviations until an reasonable selection is achieved. 
  Adjust components before adjusting number of standard deviations, increasing number incrementally until the best fit is identified.
  \end{itemize}
\end{itemize}

\subsection{Fluorescence Compensation}
\label{s:compensation}

In order to quantify the fluorescence due to the presence of a fluorescent reporter, two other source of fluorescence must be removed from its channel: cell autofluorescence (estimated from the negative control or null control), and spill-over fluorescence from other fluorescent reporters due to spectral overlap (estimated from single-color controls).
Together, this removal process is known as compensation, and is handled via linear transformations.
Here we present a minimal protocol for compensation for this context;
a more general presentation of compensation and the principles behind it may be found in~\cite{Roederer2001,Roederer2002}.

\begin{figure}
\centering
\includegraphics[width=0.6\textwidth]{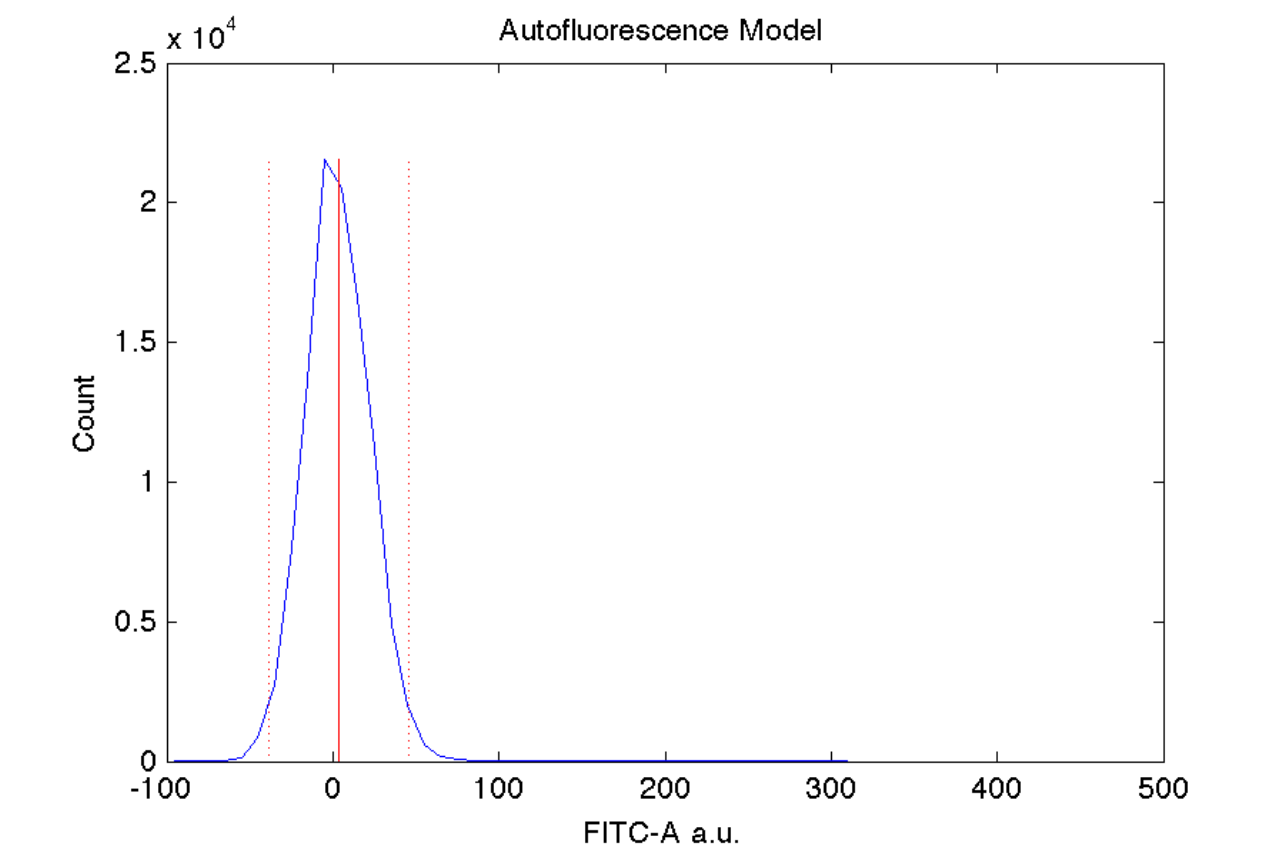}
\caption{Example of autofluorescence model computed from a negative control (solid red line shows mean, dotted red lines show $\pm 2$ std.dev.), plotted against a linear-scale histogram of events from the sample. Note the use of a linear scale and arithmetic statistics rather than logarithmic in this case. Figure adapted from~\cite{tasbe_tutorial}.}
\label{fig:autofluorescence}
\end{figure}

For each channel for a fluorescent reporter, compute an autofluorescence model for that channel:
\begin{itemize}
\item Load events from the negative control (or null transfection, if one was used per Section~\ref{sec:nt})
\item Gate events to retain only cell events, using the gate computed in Section~\ref{s:gating}.
\item Compute the arithmetic mean and standard deviation for all cell events in the sample.
  \begin{itemize}
  \item The mean will be used for compensation, subtracted from each event to convert total fluorescence to net fluorescence.
  \item The standard deviation will be used to determine where values cannot be distinguished from autofluorescence. Any value less than 2 std.dev. net fluorescence should be considered equivalent to autofluorescence. Truncating such values (e.g., to a minimum net fluorescence of 1) will also remove non-positive net fluorescence values, which are problematic for log-scale statistics.
  \item Linear statistics are used because instrument measurement noise is typically expected to dominate the variability of autofluorescence measurements. If this is not the case, it may be necessary to use a null transfections and/or to switch to geometric statistics.
  \end{itemize}
\end{itemize}

\begin{figure}
\centering
\subfigure[Single Color Control analysis]{\includegraphics[width=0.45\textwidth]{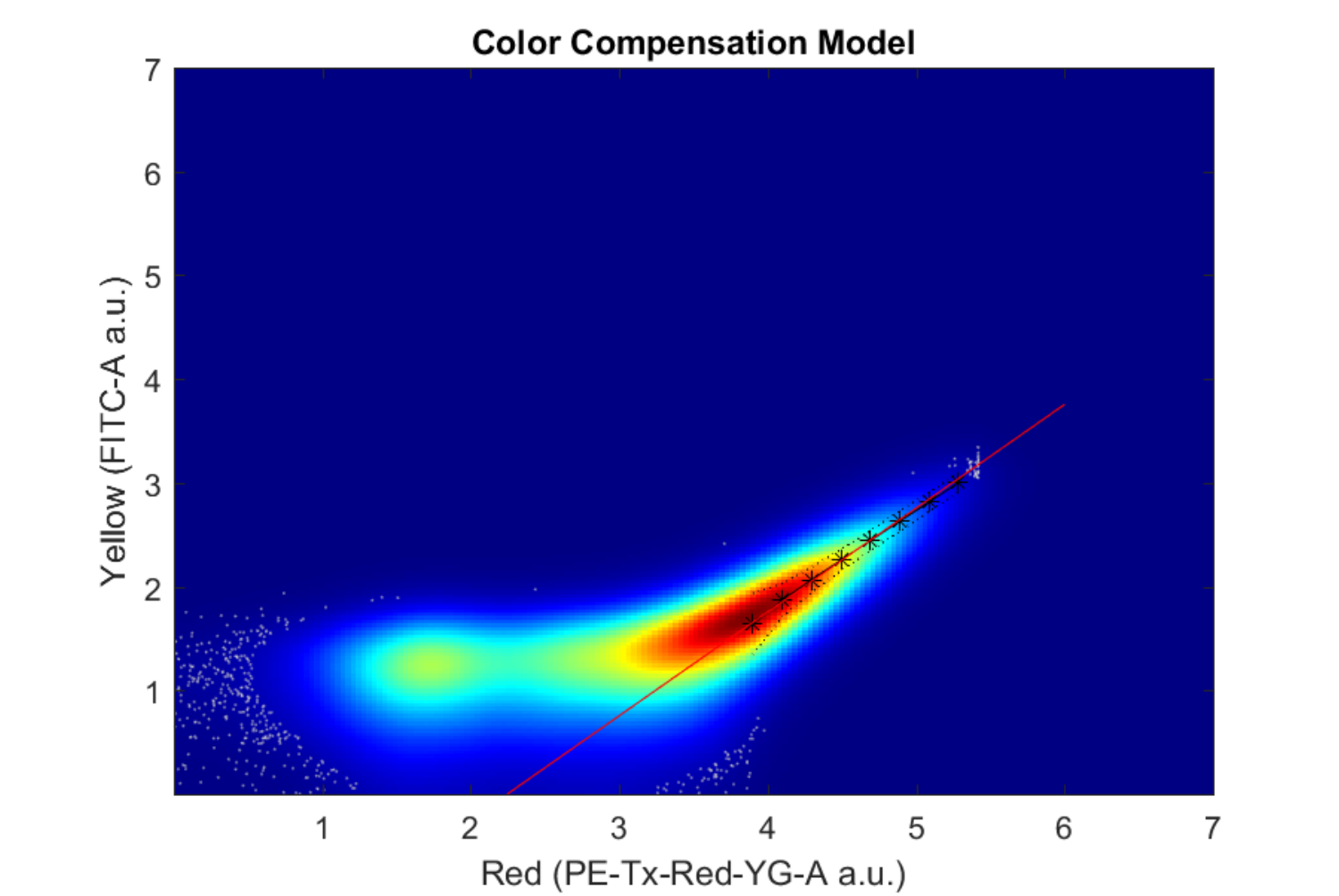}\label{f:overlap}}
\subfigure[Compensated Single Color Control]{\includegraphics[width=0.45\textwidth]{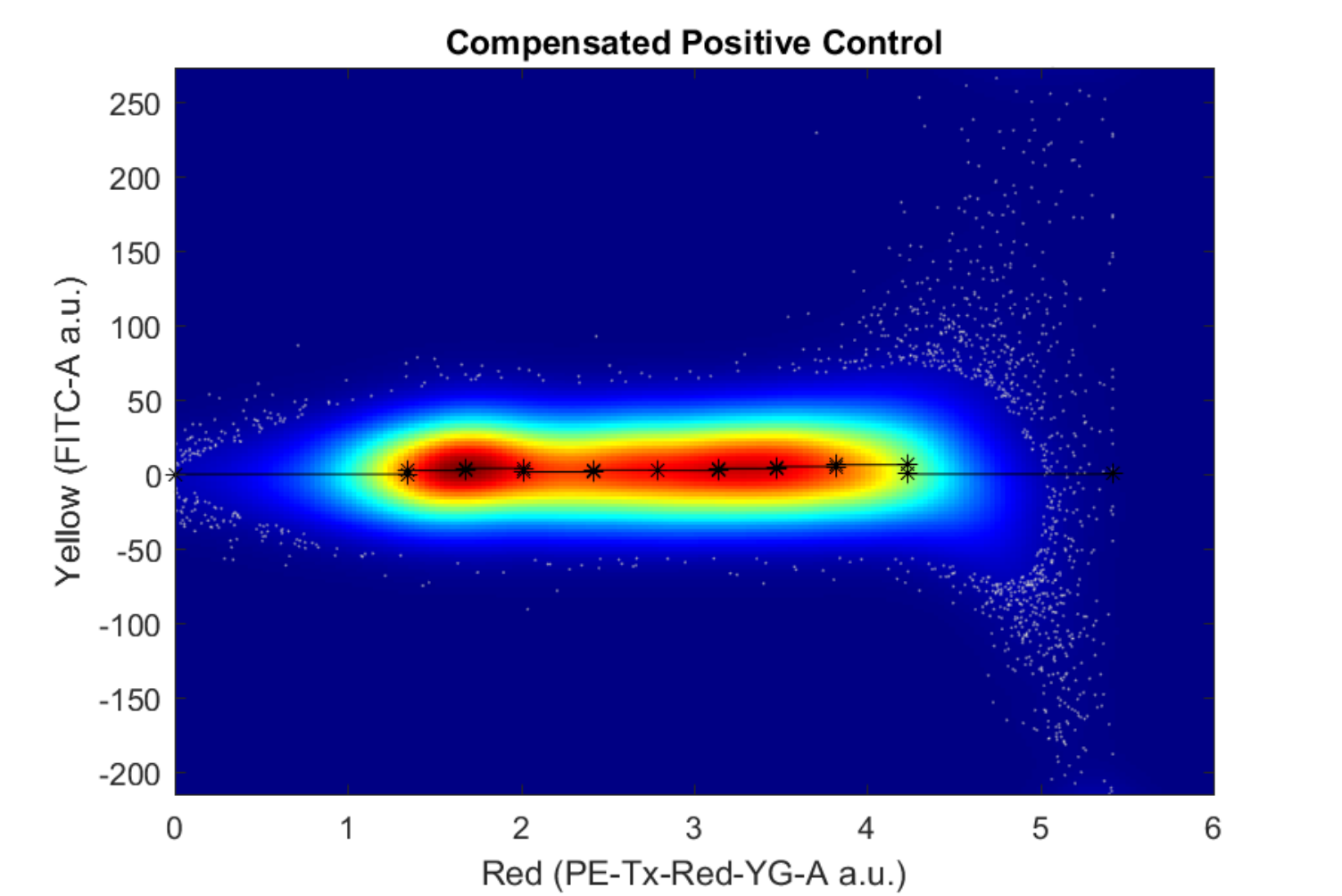}\label{f:comp_verification}}
\caption{(a) Example of calculating spectral overlap, in this case from the PE-Tx-Red-YG channel (driven) into the FITC-A channel (passive), finding a spillover rate of a little under 1\%. Black dots are the parametric means of significant bins, and the red line is the constrained linear fit for spectral overlap. (b) Example of compensation verification plot using the compensated values for the same sample and channels. Black dots are means of selected bins, and black lines are visualizing bin levels.
Figure adapted from~\cite{tasbe_tutorial}.}
\label{fig:compensation}
\end{figure}

For each channel for a fluorescent reporter, compute the spectral overlap model from that channel into other channels:
\begin{itemize}
\item Load events from the single color control for the channel.
\item Gate events to retain only cell events, using the gate computed in Section~\ref{s:gating} and subtract autofluorescence means to convert to net fluorescence.
\item Compute a parametric analysis of all other fluorescence channels (passive channels) with respect to the expressed fluorescent reporter (driven channel).
  \begin{itemize}
  \item The parametric function for each channel should either be consistently near zero (no significant overlap) or a ``dog-leg'' that is near zero for low values but then switches to a tight linear relation with unit slope at the point where spectral overlap begins to exceed autofluorescence.
  \end{itemize}
\item For each passive channel, select the significant bins by selecting only those where the passive channel mean net value is above 2 std.dev. of channel autofluorescence.
\item If there is at least one significant bin, the spectral overlap is computed as a constrained linear fit (slope=1) for passive versus driven values for the significant bins. Otherwise, spectral overlap is zero.
  \begin{itemize}
  \item For good resolution on low values, spectral overlap should be less than 10\%. 
  \item An example of computing spectral overlap is shown in Figure~\ref{f:overlap}.
  \item Compensation for spectral overlap is computed by dividing a vector of net fluorescence values by the spectral overlap matrix.
  \item Correct compensation can be verified by a log-linear plot of compensated event values for each single color control with respect to each driven channel, which should produce a balanced distribution centered on zero. An example is shown in Figure~\ref{f:comp_verification}.
  \end{itemize}
\end{itemize}

Given the autofluorescence and spectral overlap model, compensated fluorescence values are produced by first subtracting autofluorescence mean from each channel, then dividing by the spectral overlap matrix.

\subsection{Fluorescent Channel Conversion}
\label{s:conversion}

\begin{figure}
\centering
\includegraphics[width=0.6\textwidth]{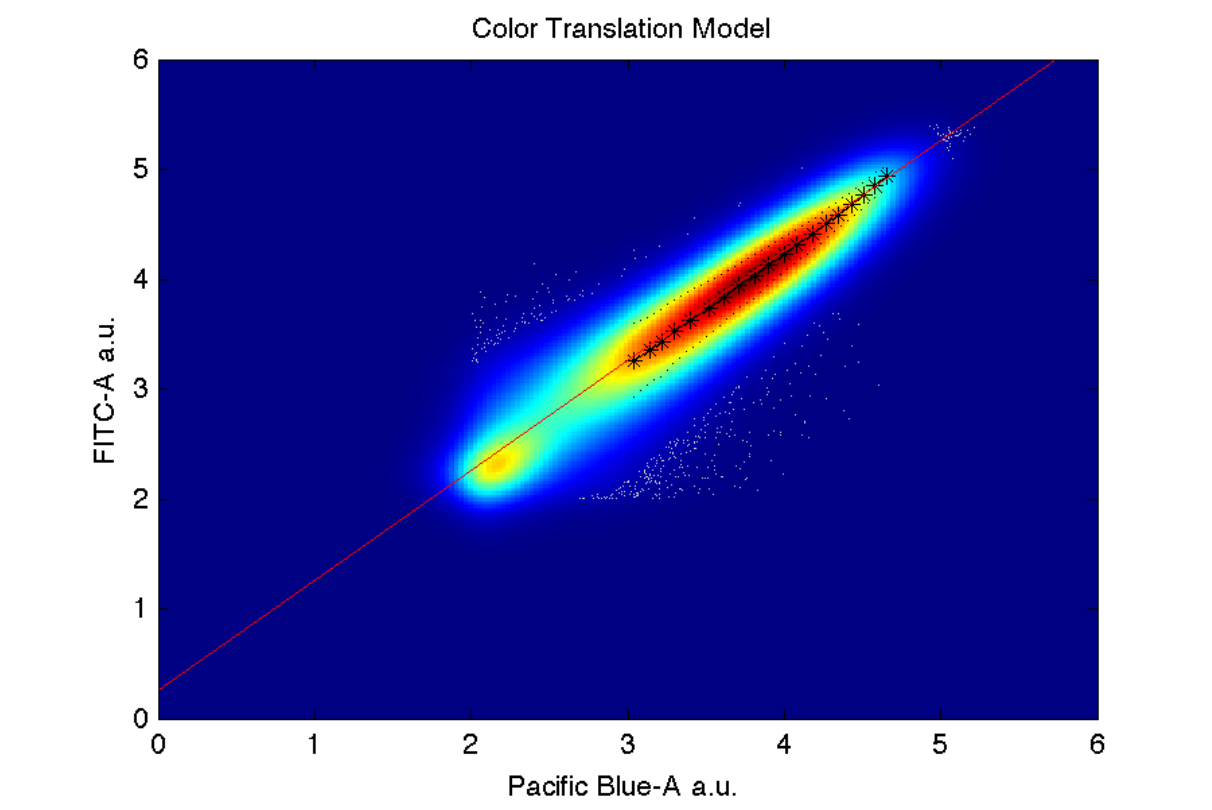}
\caption{Example of multicolor control being used to build a channel unit conversion model. Colors show density of data; black marks show parametric geometric mean of blue (Pacific Blue-A) and yellow (FITC-A) fluorescence with respect to red fluorescence (parametric axis not shown) for significant bins at 10 bins per decade. The red line shows the mean scaling factor for blue arbitrary units to yellow arbitrary units. Figure adapted from~\cite{tasbe_tutorial}.}
\label{fig:translation}
\end{figure}

\begin{itemize}
\item Load events from the multi-color control.
\item Gate events to retain only cell events, using the gate computed in Section~\ref{s:gating} and compensate to net fluorescence using the compensation model computed in Section~\ref{s:compensation}.
\item For each fluorescent channel $C$ other than the Molecules of Equivalent FLuorescein (MEFL) channel, compute a parametric analysis of  channel $C$ and the MEFL channel with respect to a third channel.
  \begin{itemize}
  \item If there are only two channels, compute a parametric analysis of the second channel against the MEFL channel.
  \item If there is no MEFL channel, another ERF-calibrated channel can be substituted, but this is not desirable because it will make it more difficult to compare data with data produced by experiments using MEFL.
  \end{itemize}
\item Select the significant bins by selecting only those where channel mean net value for both the $C$ and MEFL channels is above 2 std.dev. of channel autofluorescence.
\item The unit conversion factor from $C$ to MEFL is computed as a constrained linear fit (slope=1) for $C$ mean net values versus MEFL mean net values for the significant bins.
  \begin{itemize}
  \item Figure~\ref{fig:translation} shows an example of computing a color translation model from the Pacific Blue channel to FITC channel, parameterized by values on a Texas Red channel.
  \item If there are no significant bins or the mean fit error is more than 10\%, this indicates a problem with the multi-color control. See Section~\ref{sec:bad_translation} for likely issues.
  \end{itemize}
\end{itemize}
  
\subsection{Experimental Sample Analysis}
\label{s:expt_analysis}

\begin{figure}
\centering
\subfigure[Raw FCS data distribution]{\includegraphics[width=0.45\textwidth]{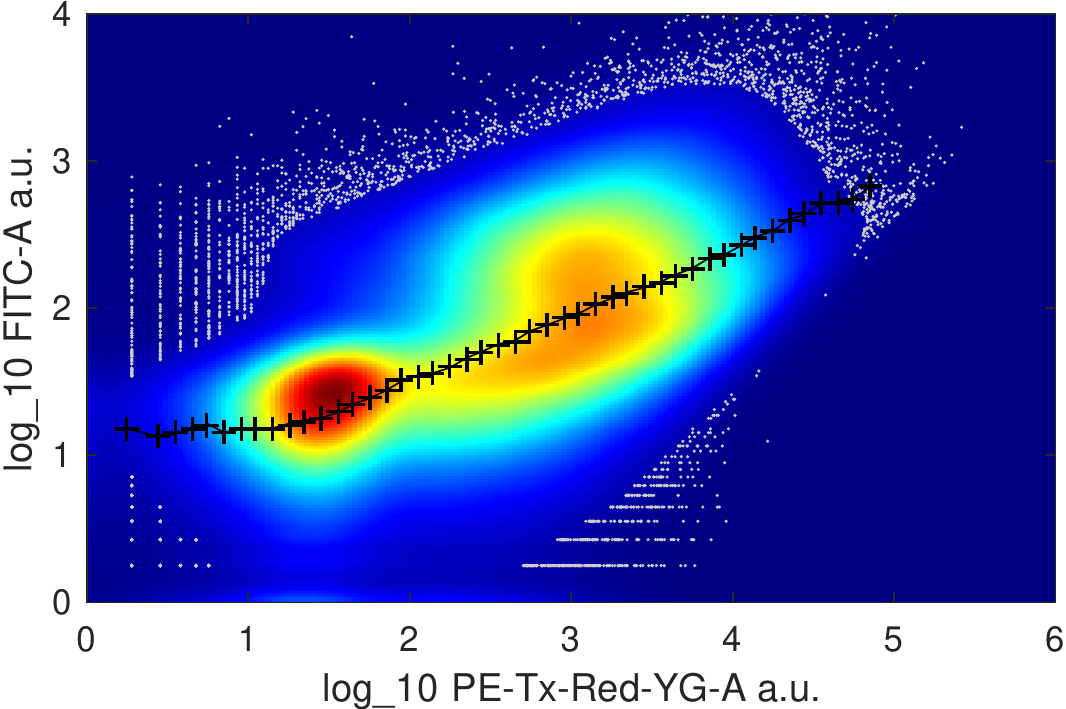}}
\subfigure[Calibrated cell data distribution]{\includegraphics[width=0.45\textwidth]{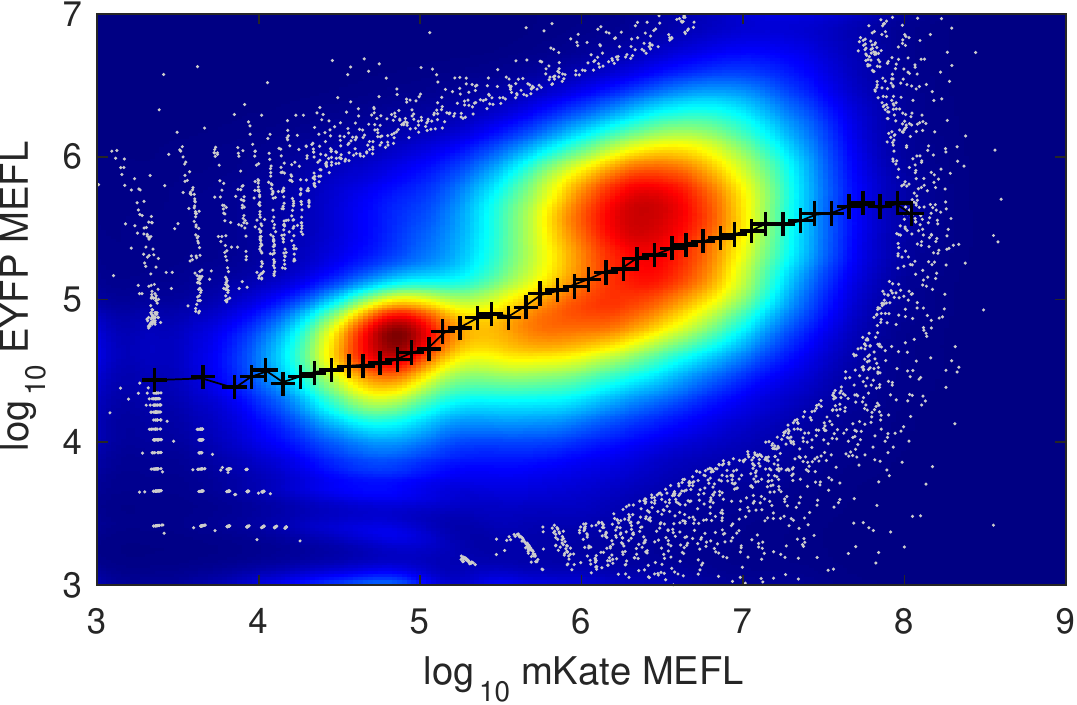}}
\caption{Example of data from a flow cytometry file transformed by the full gating and calibration model (gating, compensation color translation, ERF units). Colors show density of data; black marks show parametric geometric mean of yellow fluorescence with respect to red fluorescence at 10 bins per decade. Note that the transformed data has an enriched relative density of strongly transfected cells, that these cells have a more balanced distribution, and that the two axes have the same MEFL units. Figure adapted from~\cite{tasbe_tutorial}.}
\label{fig:color_model}
\end{figure}

\begin{itemize}
\item Load events for each experimental sample.
  \begin{itemize}
  \item Note that some of the process controls above are likely to also be used as experimental controls.
  \end{itemize}
\item Gate events to retain only cell events, using the gate computed in Section~\ref{s:gating}.
\item Compensate to net fluorescence using the compensation model computed in Section~\ref{s:compensation}.
\item Translate all fluorescence channels to the MEFL channel using the conversion factor computed in Section~\ref{s:conversion}.
\item Convert to units of MEFL (fluorescence channels) and E$\mu$m (FSC-A) using the conversion factors calculated in Section~\ref{s:colorbeads} and~\ref{s:sizebeads}, respectively.
  \begin{itemize}
  \item Figure~\ref{fig:color_model} shows an example of a sample's raw event distribution converted to a calibrated cell event distribution.
  \end{itemize}
\end{itemize}

\begin{figure}
\centering
\subfigure[Histogram Analysis]{\includegraphics[width=0.45\textwidth]{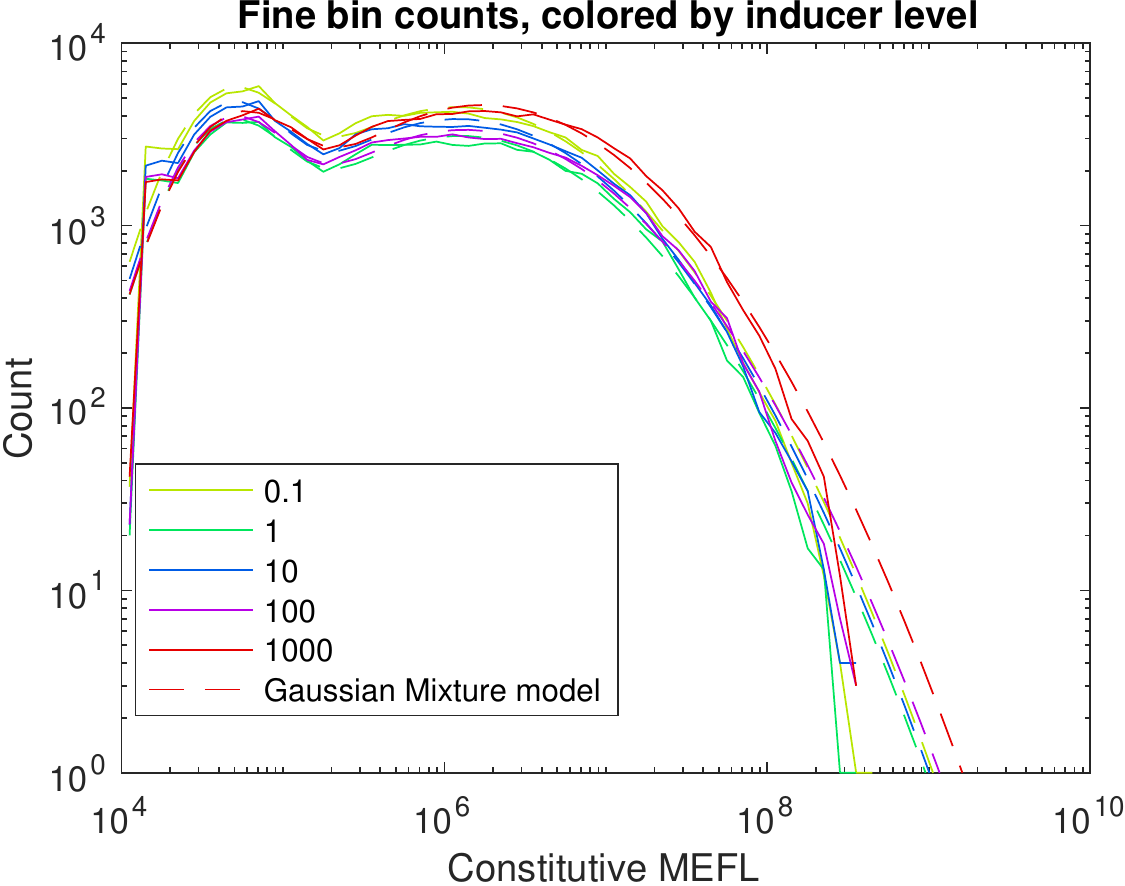}\label{f:transfection}}
\subfigure[Parametric Analysis]{\includegraphics[width=0.45\textwidth]{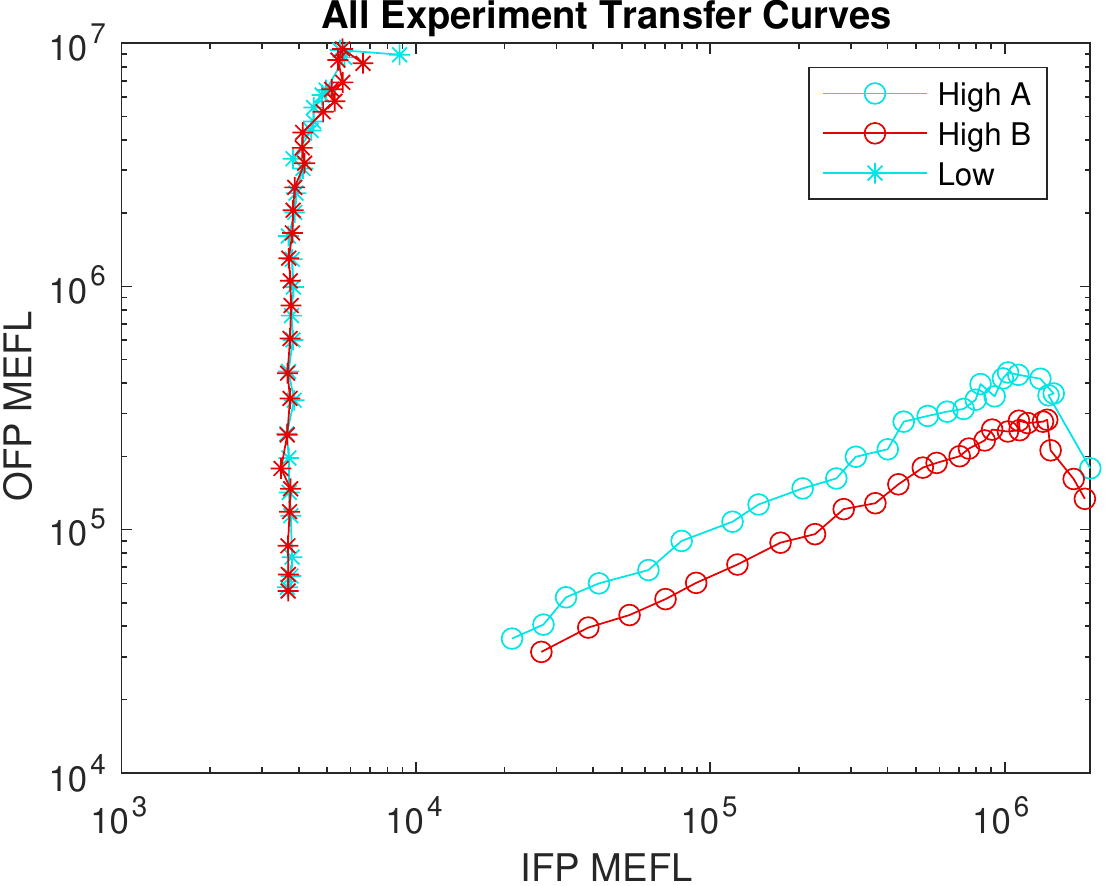}\label{f:parametric}}
\caption{Examples of log-scale statistical analysis, as applied to transient transfection samples: (a) Log-scale histogram analysis of transfection marker expression, showing a bimodal distribution of transfection levels. (b) Parametric analysis of two test conditions of a repressor under high or low induction, comparing input fluorescence protein (IFP) expression levels versus output fluorescence protein (OFP) expression levels, binned with respect to the strongly transfected range of a constitutive fluorescent protein.
Figure adapted from~\cite{tasbe_tutorial}.}
\label{fig:analyses}
\end{figure}

Once data has been gated and converted to MEFL and E$\mu$m, the further analyses that will be of value will depend on the specifics of the experiment. 
The same histogram and parametric analyses as have been used above, however, are also typically useful for analyzing some common experimental properties of interest.

\begin{itemize}
\item Distribution of transfection levels can be analyzed by fitting a Gaussian mixture model (GMM) to a log-scale histogram analysis of a constitutive fluorescent reporter.
  \begin{itemize}
  \item Figure~\ref{f:transfection} shows a typical transient transfection distribution from lipofection. 
  \item There are typically two major components: strongly transfected cells and transfection failures. A strong lipofection should typically have more than 50\% of the cells in the strongly transfected component.
  \item If less than 30\% are in the strongly transfected component, this usually indicates a failure of the transfection protocol.
  \item Biologically, it is not clear whether transfection failure means no transfection or just few copies. The transfection failure component should typically be excluded from experimental analysis in any case, as the behavior of cells in the transfection failure component are generally qualitatively different from those in the strongly transfected component.
  \item In some cases, the fit is better when a third component is included in the middle, though it is unclear whether this represents a true third population or if the better fit is due to a distribution asymmetry, e.g., due to cellular resource limitations.
  \item A significant drop in the strongly transfected component from run to run or from sample to sample is often an indication of problems in protocol execution that will render the data from the less-well-transfected samples unreliable.
  \end{itemize}
\item The behavior of transfected constructs can be analyzed by a parametric analysis of the experimental fluorescent reporters with respect to a constitutive fluorescent reporter.
  \begin{itemize}
  \item With transient transfection, parametric analysis generally provides more insight than bulk statistical analysis because of the high variability in transfection level.
  \item The parametric range should be limited to only strongly transfected cells.
  \item Figure~\ref{f:parametric} shows an example of comparing test conditions using parametric analysis with respect to a constitutive fluorescent reporter.
  \end{itemize}
\end{itemize}

\section{Notes}
\label{sec:notes}

This section presents common issues that can occur in quantifying transient transfections in mammalian cells with flow cytometry and adjustments to the method that can be used to resolve these issues.

\subsection{Null Transfection Control}
\label{sec:nt}

Some strains of cell respond to some transfection protocols or some vector backbones with large increases in autofluorescence. 
This is most often in red channels, and often indicates that cell viability is degraded.
If this is suspected to be the case, then a null transfection control should be added, in which cells are transfected with a vector that contains none of the constructs under study.
This construct will then be used for estimating autofluorescence instead of the wild-type, but wild-type will still be used for assessing strain response to culturing.

\subsection{Small fraction of cell-like particles}

If the number of cell-like particles making it through gating is small (under 50\%), this may indicate problems with data collection. This may be diagnosed with a plot of forward versus side scatter for the wild type negative control.
\begin{itemize}
\item If the distribution of cell-like particles is truncated against a low threshold, the instrument's forward scatter trigger for event capture likely needs to be lowered.
\item If there are more smaller particles than cell-like particles, the instrument's forward scatter trigger for event capture likely needs to be raised.
\item If shifting triggers does not correct the problem, then the flow rate may need to be lowered in order to decrease the rate at which events occurs.
\end{itemize}

\subsection{Alternative Statistics Analyses}
\label{s:alt_stats}

Many flow cytometry analyses recommend the use of median or mode values instead of geometric means.
The median often gives values that are very similar to the geometric mean, but it does not have a biological distribution theory supporting its use. As a result, although other percentiles can provide some notion of deviation, there is no principled equivalent of geometric standard deviation of multi-modal log-normal fits to guide the interpretation of median-based analyses.

Mode-based analysis is typically presented as finding a peak or maximum in a histogram. 
While this works well in some cases, biological variability and the inherent quantization associated with binning to produce a histogram means that there is often significant uncertainty in the location of the mode, and the value can be strongly affected by the details of how a histogram is computed. 
The same qualitative goals can thus often be better satisfied by use of multi-modal geometric statistics.

Finally, it is useful to note that log-scale statistics are ill-defined for non-positive numbers and thus do not provide correct results for small values, where error tends to be dominated by arithmetic contributions from instrument noise or fluorescence compensation.
This is often addressed in visualization by using a bi-exponential or ``logicle'' remapping~\cite{tung2007modern}, which provides a clean visual transition between regions of geometric and arithmetic error dominance, but which involves an arbitrary cutoff and does not shed light on the underlying question of biological activity. 
Better methods for incorporation of low and negative flow cytometry values into analysis remains an open question.

\subsection{Bead Peak Problems}
\label{s:bead_problems}

In addition to the notes already presented in Section~\ref{s:colorbeads} and~\ref{s:sizebeads}, other common issues in identifying bead peaks are small counts, high background, blurred peaks, and offset peaks.
\begin{itemize}
\item If the number of events per peak is small, on the order of 200 or less per peak, then the precision of unit conversion is likely to be poor. This issue also typically indicates problems with the FSC threshold for event capture.
  \begin{itemize}
  \item If the total number of events in the bead file is low, this typically indicates that the event capture threshold is too high. 
  \item If the total number of events is as expected, but most are low values not forming clean peaks, this typically indicates that the event capture threshold is too low.
  \end{itemize} 
\item High levels of background events between peaks can make peaks difficult to identify or separate. If this occurs for values higher than around the $10^2$ arbitrary units, this typically indicates an instrument that is dirty or has other service issues that are creating the high rates of background events.
\item If there are peaks, but they do not generally have a single highly distinct maximum (e.g., wide peaks or doubled peaks), then this generally indicates an instrument hardware issue that requires service, such as lasers having come out of alignment.
\item If there are clear peaks, but they do not produce a good linear fit, this typically indicates that the peak identities have been offset from their true identities, e.g., peaks 4-6 have been mis-labelled as peaks 2-4.
\end{itemize}

\subsection{Compensation Problems}
\label{s:bad_compensation}

\begin{itemize}
\item If spectral overlap is more than 10\%, the level of effective overlap can be reduced by adjusting choice of fluorescent reporter and/or channel voltage. 

\item Expression of the single color control is too weak if it could not be used to identify at least a 1\% spectral overlap. Typically this means that there should be significant numbers of events through at least the $10^4$ to $10^5$ arbitrary unit range. Weak expression can be caused by a number of issues:
  \begin{itemize}
  \item The selected channel may not be a close enough match for the excitation and emission properties of the fluorescent reporter, such that only a small fraction of the potential fluorescent energy is being measured, in which case either the channel or the reporter should be changed.
  \item The fluorescent reporter may have poor brightness or there may be stronger than usual  autofluorescence in the selected channel, in which case the reporter should be changed.
  \item The fluorescent reporter may be a poor choice for the protocol (e.g., mismatched chemical properties), in which case the reporter should be changed.
  \item If none of these is the case, then it is an indicator that there are problems with the reporter construct (e.g., poor promoter strength, interference between promoter and reporter) and/or the experimental protocol.
  \end{itemize}

\item If compensation verification shows a failure, i.e., the center of the distribution does not stay roughly centered on zero, this is typically caused by problems with one or more of the single color controls or their analysis. 
  \begin{itemize}
  \item Weak expression can cause compensation failures, and should be addressed as discussed above.
  \item If the control has a ``messy'' distribution instead of the expected pattern of autofluorescence followed by a tight linear unit slope, then it typically indicates problems in either the reporter construct, experimental protocol, or cell gating.
  \item Instrument problems can also cause compensation failures, but it is highly unlikely for this to be the case without the problem already being identified during peak identification for the color beads.
  \end{itemize}
\end{itemize}

\subsection{Channel Conversion Problems}
\label{sec:bad_translation}

\begin{itemize}
\item If there are no significant bins or there is a poor fit (more than 10\% error), first check the single color controls to see if their expression is weak or ``messy,'' per Section~\ref{s:bad_compensation}.
\item If all of the single color controls are strong and have clean distributions, then it is likely that there is a problem with the co-transfection protocol or the specifics of the multi-color control. 
\end{itemize}

\subsection{Transfection Problems}
\label{s:transfection_problems}

\begin{figure}
\centering
\subfigure[Problematic Samples]{\includegraphics[width=0.45\textwidth]{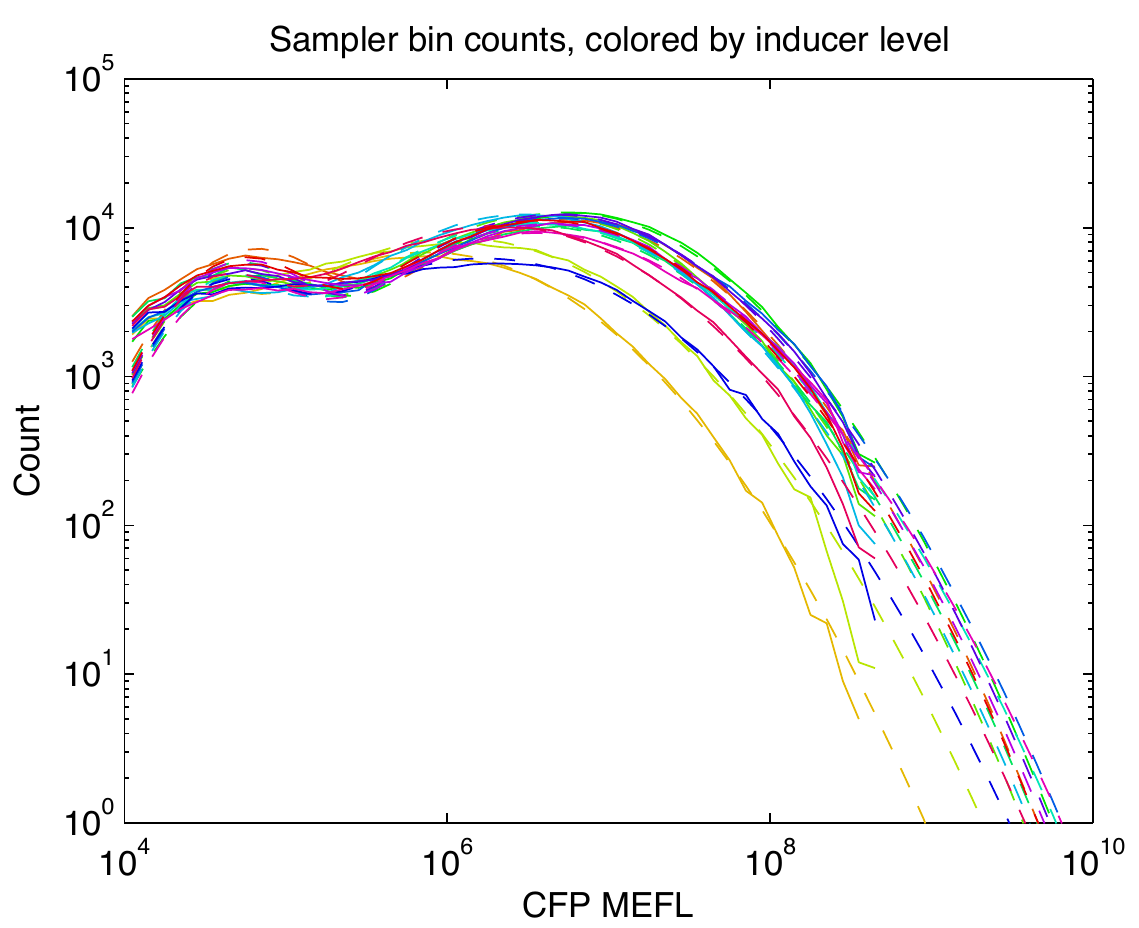}\label{f:bad_samples}}
\subfigure[Problematic Experiment]{\includegraphics[width=0.45\textwidth]{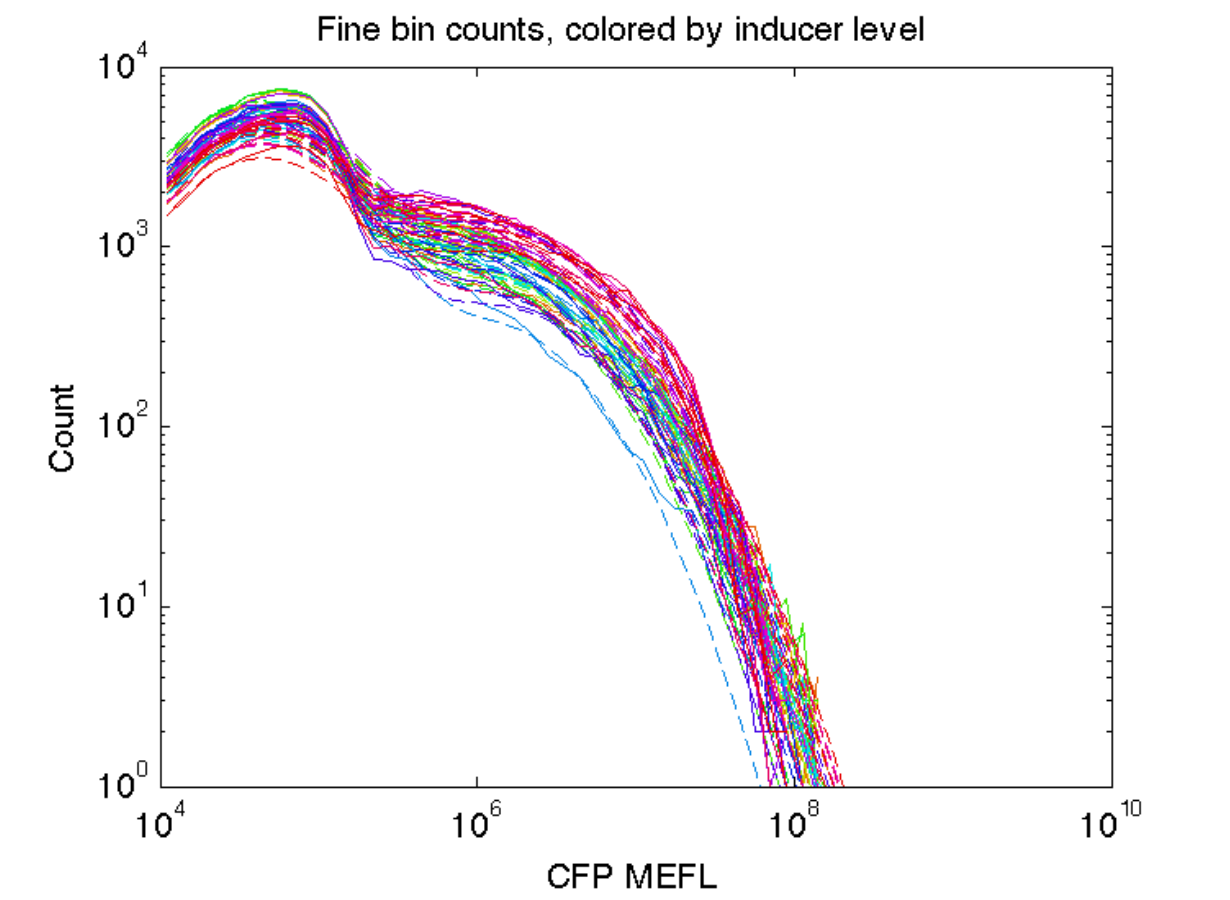}\label{f:bad_batch}}
\caption{Example of distribution expectations allowing identification of individual bad samples (a) and a full batch of samples gone wrong from a badly executed protocol (b).}
\label{fig:bad_transfection}
\end{figure}

Figure~\ref{fig:bad_transfection} shows examples of experimental samples failing quality control due to differences between expected and observed transfection distributions.
In Figure~\ref{f:bad_samples}, the lowest three samples have significantly lower strongly transfected components than the rest, which are tightly bunched. Those three samples were thus considered to be likely protocol failures and excluded, even though they have more than 50\% of their events in the strongly transfected component.
In Figure~\ref{f:bad_batch}, the entire collection of samples is showing a strongly transfected component containing less than 30\% of the cell events, and thus the entire experimental run was considered likely to have been affected by a protocol failure and was re-run.

\begin{acknowledgement}
This document does not contain technology or technical data controlled under either the U.S. International Traffic in Arms Regulations or the U.S. Export Administration Regulations.
\end{acknowledgement}

\bibliographystyle{plain}
\bibliography{references}

\end{document}